\documentclass[runningheads,a4paper]{llncs}
\usepackage{amssymb}
\usepackage{graphicx}
\usepackage{url}
\newcommand{\keywords}[1]{\par\addvspace\baselineskip
\noindent\keywordname\enspace\ignorespaces#1}
\begin{document}

\mainmatter


\newtheorem{Example}{\bf Example}
\newtheorem{Remark}{\bf Remark}
\newtheorem{Table}{\bf Table}
\newtheorem{Notation}{\bf Notation}

\def\T {\ensuremath{\bf{T}}}
\def\U {\ensuremath{\bf{U}}}
\def\N {\ensuremath{\mathbb{N}}}
\def\C {\ensuremath{\mathbb{C}}}
\def\R {\ensuremath{\mathbb{R}}}
\def\Q {\ensuremath{\mathbb{Q}}}
\def\A {\ensuremath{\bf{A}}}
\def\B {\ensuremath{\bf{B}}}
\def\P {\ensuremath{\bf{P}}}
\def\S {\ensuremath{\mathbb{S}}}
\def\E {\ensuremath{\bf{E}}}
\def\H {\ensuremath{\bf{H}}}
\def\V {\ensuremath{\rm{V}}}
\def\D {\ensuremath{\rm{D}}}
\def\PF {\ensuremath{\bf {PF}}}
\def\TH {\ensuremath{\bf{TH}}}
\def\RS {\ensuremath{\mathbb{T}}}
\def\HCTD {\ensuremath{\tt{HCTD}}}
\def\HPCTD {\ensuremath{\mathrm{HPCTD}}}
\def\CTD {\ensuremath{\mathrm{CTD}}}
\def\PCTD {\ensuremath{\mathrm{PCTD}}}
\def\WUCTD {\ensuremath{\mathrm{WUCTD}}}
\def\RSD {\ensuremath{\mathrm{RSD}}}
\def\WCTD {\ensuremath{\mathrm{WCTD}}}
\def\FWCTD {\ensuremath{\mathrm{FWCTD}}}
\def\SWCTD {\ensuremath{\mathrm{SWCTD}}}
\def\ARSD {\ensuremath{\tt{RSD}}}
\def\APRSD {\ensuremath{\tt{WRSD}}}
\def\CCTD {\ensuremath{\tt{CTD}}}
\def\ASWCTD {\ensuremath{\tt{SWCTD}}}
\def\ASMPD {\ensuremath{\tt{SMPD}}}
\def\AHPCTD {\ensuremath{\tt{HPCTD}}}
\def\TDU {\ensuremath{\mathrm{TDU}}}
\def\ATDU {\ensuremath{\tt{TDU}}}
\def\RDU {\ensuremath{\mathrm{RDU}}}
\def\ARDU {\ensuremath{\tt{RDU}}}
\newtheorem{Rules}{\bf Rule}

\newcommand{\disc}[1]{\mbox{{\rm disc}$(#1)$}}
\newcommand{\alg}[1]{\mbox{{\rm alg}$(#1)$}}
\newcommand{\SAT}[1]{\mbox{{\rm SAT}$(#1)$}}
\newcommand{\ideal}[1]{\langle#1\rangle}
\newcommand{\I}[1]{\mbox{{\rm I}$_{#1}$}}
\newcommand{\ldeg}[1]{\mbox{{\rm ldeg}$(#1)$}}
\newcommand{\iter}[1]{\mbox{{\rm iter}$(#1)$}}
\newcommand{\mdeg}[1]{\mbox{{\rm mdeg}$(#1)$}}
\newcommand{\lv}[1]{\mbox{{\rm lv}$_{#1}$}}
\newcommand{\mvar}[1]{\mbox{{\rm mvar}$(#1)$}}
\newcommand{\red}[1]{\mbox{{\rm red}$(#1)$}}
\newcommand{\prem}[1]{\mbox{{\rm prem}$(#1)$}}
\newcommand{\pquo}[1]{\mbox{{\rm pquo}$(#1)$}}
\newcommand{\rank}[1]{\mbox{{\rm rank}$(#1)$}}
\newcommand{\res}[1]{\mbox{{\rm res}$(#1)$}}
\newcommand{\cls}[1]{\mbox{{\rm cls}$_{#1}$}}
\newcommand{\sat}[1]{\mbox{{\rm sat}$(#1)$}}
\newcommand{\sep}[1]{\mbox{{\rm sep}$(#1)$}}
\newcommand{\tail}[1]{\mbox{{\rm tail}$(#1)$}}
\newcommand{\zm}[1]{\mbox{{\rm MZero}$(#1)$}}
\newcommand{\zero}[1]{\mbox{{\rm Zero}$(#1)$}}
\newcommand{\rd}[1]{\mbox{{\rm Red}$(#1)$}}
\newcommand{\map}[1]{\mbox{{\rm map}$(#1)$}}

 \newcommand {\bbb}[1]{\raisebox{-1.5ex}[0pt][0pt]{\shortstack{#1}}}

\title{Hierarchical Comprehensive Triangular Decomposition\thanks{The work was supported by National Science Foundation of China (Grants
11290141 and 11271034).
}}  
\titlerunning{Hierarchical CTD} 
\author{Zhenghong Chen\inst{} \and  Xiaoxian Tang\inst{} \and Bican Xia\inst{} }
\authorrunning{Chen-Tang-Xia}
\institute{School of Mathematical Sciences,
Peking University, China\\
\email{septemwnid@pku.edu.cn, tangxiaoxian@pku.edu.cn, xbc@math.pku.edu.cn}\\
}
\maketitle

\begin{abstract}
The concept of {\em comprehensive triangular decomposition} (CTD) was first introduced by Chen {\it et al.} in their CASC'2007 paper and could be viewed as an analogue of comprehensive Gr\"obner systems for parametric polynomial systems.
The first complete algorithm for computing CTD was also proposed in that paper and implemented in
the {\tt RegularChains} library in Maple. Following our previous work on generic regular decomposition for parametric polynomial systems, we introduce in this paper a so-called {\em hierarchical} strategy for computing CTDs. Roughly speaking, for a given parametric system, the parametric space is divided into several sub-spaces of different dimensions and we compute CTDs over those sub-spaces one by one. So, it is possible that, for some benchmarks, it is difficult to compute CTDs in reasonable time while this strategy can obtain some ``partial" solutions over some parametric sub-spaces.
The program based on this strategy has been tested on a number of benchmarks from the literature. Experimental results on these benchmarks with comparison to {\tt RegularChains} are reported and may be valuable for developing more efficient triangularization tools.
\end{abstract}

\section{Introduction}\label{Intro}
 Solving parametric polynomial system plays a key role in many areas such as automated geometry theorem deduction,  stability analysis of dynamical systems, robotics and so on. For an arbitrary parametric system, in symbolic computation, solving this system is to convert equivalently the parametric system into new systems with special structures so that it is easier to analyze or solve the solutions to the new systems. There are two main kinds of symbolic methods to solve parametric systems, {\it i.e.}, the algorithms based on {\em Gr\"obner bases} \cite{sun,Montes,KN,SS,newSS,CGS} and those based on {\em triangular decompositions} \cite{ap,marco,changbo,gxs1992,kalk,maza,dkwang,wangi,wu,yhx01,xia,zjzi}.

The methods based on triangular decompositions have been studied by many researchers since Wu's work \cite{wu} on {\em characteristic sets}. A significant concept in the theories of triangular sets is {\em regular chain} (or {\em normal chain}) introduced by Yang and Zhang \cite{zjzi} and  Kalkbrener \cite{kalk} independently. Gao and Chou proposed a method in \cite{gxs1992} for identifying all parametric values for which a given system has solutions and giving the solutions by $p-${\em chains}
without a partition of the parameter space.
Wang generalized the concept of regular chain to {\em regular system} and gave an efficient algorithm for computing it \cite{wangi,wang,wangEpsilon}. It should be noticed that, due to their strong projection property, the regular systems or series 
may also be used as representations for parametric systems. Chen {\it et. al.} introduced the concept of {\em comprehensive triangular decomposition} (CTD) \cite{changbo} to solve parametric systems, which could be viewed as an analogue of comprehensive Gr\"obner systems. Algorithm {\tt CTD} for computing CTD was also proposed in \cite{changbo}.

There are several implementations based on the above triangularization methods, such as {\tt Epsilon} \cite{wangEpsilon}, {\tt RegularChains} \cite{changbo2011} and {\tt wsolve} \cite{dkwang}.

Suppose ${\P}\subset \Q[U][X]$ is a parametric polynomial system where $X=(x_1,\ldots,\allowbreak x_n)$ are variables and $U=(u_1,\ldots,u_d)$ are parameters. The above mentioned algorithms all solve the system in $\C^{d+n}$ directly. That means all the unknowns ($U$ and $X$) are viewed as variables and triangular decompositions are computed over $\Q$. It may happen that no triangular decompositions over $\Q$ can be obtained in a reasonable time for some systems while a triangular decomposition over $\Q[U]$ is much easier to be computed.

Based on this observation, we propose a strategy which  computes CTDs for given parametric systems hierarchically and the CTDs are called {\em hierarchical comprehensive triangular decompositions} (HCTD). By ``hierarchical" we mean that, roughly speaking, a generic regular decomposition is computed first over $\Q[U]$ and a parametric polynomial $B(U)$ is obtained at the same time such that the solutions to the original system in $\C^{d+n}$ can be expressed as the union of solutions to those regular systems in the decomposition provided that the parameter values satisfying $B(U)\ne 0$. Then, by applying similar procedure recursively, the solutions satisfying $B(U)=0$ are obtained through adding $B(U)=0$ to the system and treating some parameters as variables. We give an algorithm based on this hierarchical strategy which computes CTDs for given parametric systems. The algorithm has been implemented with Maple and  tested on a number of benchmarks from the literature. Experimental results on these benchmarks with comparison to {\tt RegularChains} are reported (see Tables \ref{res1}) and may be valuable for developing more efficient triangularization tools.
For some benchmarks, it is difficult to compute CTDs in reasonable time while our program can output ``partial solutions" (see Table \ref{res4}).

The rest part of this extended abstract is organized as follows. Section \ref{smainidea} introduces an algorithm, Algorithm {\tt HCTD}, for computing CTDs hierarchically and an example is illustrated there.
Section \ref{scwc} compares the Algorithm {\tt HCTD} and the Algrotihm {\tt CTD} in \cite{changbo} by experiments. Section \ref{sva} introduces another hierarchical strategy for computing CTD and the comparing experiments are also shown.
Section \ref{sps} shows the benefit of the hierarchical strategy by experiments. 


 \section{Algorithm {\tt HCTD}}\label{smainidea}

For the concepts and notations without definitions, please see
  \cite{marco,wang,changbo2011}.

Suppose ${\T}$ is a regular chain in  ${\mathbb Q}[U][X]$
 and  ${\bf H} \subset {\mathbb Q}[U][X]$. $[{\T},{\bf H}]$ is said to be a {\em regular system} \cite{changbo} if ${\rm res}(H, {\T}) \neq 0$ for  any $H \in {\bf H}$.
For any ${\bf B}\subset {\mathbb Q}[U]$,  ${\rm V}^{U}({\bf B})$ denotes the set
$\{(a_1,  \ldots,  a_d)\in {\mathbb C}^{d}|B(a_1,  \ldots,  a_d)=0,  \forall B\in \B\}.$
For any ${\bf P}\subset {\mathbb C}[X]$,  ${\rm V}({\bf P})$ denotes the set
$\{(b_1,  \ldots,  b_n)\in {\mathbb C}^{n}|P(b_1,  \ldots,  b_n)=0,  \forall P\in \P\}.$
For any ${\bf P}\subset {\mathbb Q}[U][X]$,  ${\rm V}({\bf P})$ denotes the set
$\{(a_1,\ldots,a_d, b_1,  \ldots,  b_n)\in {\mathbb C}^{d+n}|P(a_1,  \ldots,  a_d, b_1,\ldots,b_n)=0,  \forall P\in \P\}.$
For $D\subset {\mathbb C}^{d+n}$, denote by $\Pi_{U}(D)$ the set
$\{(a_1,\ldots,a_d)\in {\mathbb C}^{d}|(a_1,  \ldots,  a_d, b_1,\ldots,b_n)\in D\}$.
 Suppose $[{\T}, {\bf H}]$ is a regular system in ${\mathbb Q}[U][X]$. If ${\bf H}=\{H\}$, then $[{\T}, {\bf H}]$
 is denoted by $[{\T}, H]$ for short.

Due to page limitation, we only present the specification of an algorithm for computing CTDs hierarchically.
\vspace{0.3cm}

 \textbf{Algorithm HCTD}

 \textbf{Input}: a finite set ${\P}\subset {\mathbb Q}[U][X]$,  a non-negative  integer $m$ $(0\leq m\leq d)$

\textbf{output}: \begin{minipage}[t]{100mm}
    finitely  many $3$-tuples $[{\bf A}_i, {\bf B}_i, \mathbb{T}_{i}]$,  a polynomial $B$, where
     \begin{itemize}
     \item $B\in {\mathbb Q}[u_{m+1},\ldots,u_d]$, ${\bf A}_i, {\bf B}_i\subset {\mathbb Q}[U]$
     \item $\mathbb{T}_{i}$ is a finite set of regular systems in ${\mathbb Q}[U][X]$
     \end{itemize}
     such that
      \begin{itemize}
     \item $\cup_{i}{\V}^{U}({\bf A}_i\backslash {\bf B}_i)=\left({\mathbb C}^d\backslash {\V}^{U}(B)\right)\cap \Pi_{U}\left({\V}({\P})\right)$
     \item for any $i,j\;(i\neq j)$, ${\V}^{U}({\bf A}_i\backslash {\bf B}_i)\cap {\V}^{U}({\bf A}_j\backslash {\bf B}_j)=\emptyset$
    \item for any $i$, if $a\in {\V}^{U}({\bf A}_i\backslash {\bf B}_i)$, then  $[{\T}(a), {\bf H}(a)]$ is a regular system
     in ${\mathbb C}[X]$ for any $[{\bf T}, {\bf H}]\in \mathbb{T}_i$
     \item for any $i$, if  $a\in {\V}^{U}({\bf A}_i\backslash {\bf B}_i)$, then\\
     ${\V}({\P}(a))=\cup_{[{\T}, {\bf H}] \in \mathbb{T}_i} {\V}({\T}(a)\backslash {\bf H}(a)).$
     \end{itemize}
     \end{minipage}
\vspace{0.3cm}

The output of ${\tt HCTD}({\P}, m)$ is called the $m$-HCTD of $\P$. Each $[{\bf A}_i, {\bf B}_i, {\mathbb T}_i]$ in the $m$-HCTD is called a {\tt branch}. Each regular system in the set $\cup{\mathbb T}_i$ is called a {\tt grape}.
By Algorithm {\tt HCTD},  for any ${\P}$,
if $m=0$, the output is the so-called {\em generic regular decomposition \cite{hong}} of ${\P}$;
if $m=d$, the output is the {\em comprehensive triangular decomposition \cite{changbo}} of ${\P}$.
The Example \ref{ex1} below shows how to get $m$-HCTD $(m=0,\ldots,d)$.

\begin{Example}\label{ex1}
Consider the parametric system
{\footnotesize
\[{\P}=\left\{\begin{array}{l}
2x_2^2(x_2^2+x_1^2)+(u_2^2-3u_1^2)x_2^2-2u_2x_2^2(x_2+x_1)+2u_1^2u_2(x_2+x_1)\\
~~~~ -u_1^2x_1^2+u_1^2(u_1^2-u_2^2),\\
4x_2^3+4x_2(x_2^2+x_1^2)-2u_2x_2^2-4u_2x_2(x_2+x_1)+2(u_2^2-3u_1^2)x_2+2u_1^2u_2,\\
4x_1x_2^2-2u_2x_2^2-2u_1^2x_1+2u_1^2u_2.
\end{array}\right.
\]}
where $x_1$, $x_2$ are variables and $u_1$, $u_2$ are  parameters. 

   1.   By the Algorithm {\tt RDU} in \cite{hong},  we compute a set  $\mathbb{T}_{1}$ of regular systems
   and a polynomial $B_1(u_1, u_2)$ such that
   if $B_1(u_1, u_2) \neq 0$, then the solution set of ${\P}=0$ is equal to the union of the solution sets of the regular systems in $\mathbb{T}_{1}$.
    Then we obtain the $0$-HCTD  of ${\P}$:
                                         $[{\bf A}_1, {\bf B}_1, {\mathbb T}_1].$

   2. Let ${\P}_1= {\P} \cup \{ B_1\}$. Regard $\{u_1, x_1, x_2\}$ as the new variable set.  By the Algorithm {\tt RDU},
   we compute a set $\mathbb{S}_1$ of regular systems
   and a polynomial $B_2(u_2)$ such that if $B_1(u_1, u_2)=0$ and $B_2(u_2)\neq 0$, then the solution set of ${\P}=0$ is equal to the union of the solution
   sets of the regular systems in $\mathbb{S}_1$.
   For ${\mathbb S}_1$, applying the similar method as the Algorithm {\tt RegSer} in \cite{wangi} and the Algorithms {\tt Difference} and {\tt CTD} in \cite{changbo2011},  we obtain
   the $1$-HCTD  of ${\P}$: $[{\bf A}_1, {\bf B}_1, {\mathbb T}_1],\ldots,[{\bf A}_4, {\bf B}_4, {\mathbb T}_4].$

   3. Let ${\P}_2= {\P}_1 \cup \{ B_2\}$.  Regard $\{u_2, u_1, x_1, x_2\}$ as the new variable set.
    By the Algorithm {\tt RDU}, we compute a set of regular systems $\mathbb{S}_2$
    and a polynomial $B_3=1$ such that if $B_1(u_1,u_2)=0, B_2(u_2)=0$ and $B_3\neq 0$, then
      the solution set of ${\P}=0$ is equal to the union of the solution sets of the regular systems in $\mathbb{S}_2$.
    For ${\mathbb S}_2$,   applying the similar method as the Algorithms {\tt RegSer}, {\tt Difference} and {\tt CTD},  we obtain
    the $2$-HCTD  of ${\P}$:
    $[{\bf A}_1, {\bf B}_1, {\mathbb T}_1],\ldots,[{\bf A}_6, {\bf B}_6, {\mathbb T}_6].$
   \begin{Table}\label{res55}\footnotesize
  \begin{center}
       {$[{\bf A}_i, {\bf B}_i, {\mathbb T}_i]$}\\
       \begin{tabular}{c|c|c|c}
       \hline
        $\;\;i\;\;$ & ${\bf A}_i$ & ${\bf B}_i$ & $\mathbb{T}_i$ \\
       \hline
        $\;\;1\;\;$ &$\emptyset$ &  $\{u_1u_2(u_1^2-u_2^2) \}$ & $\{ [\{-2x_1^2+3x_1u_2-u_2^2+u_1^2, 2x_1x_2+u_1^2-u_2x_2 \},u_1] \}$ \\
        \hline
        $\;\;2\;\;$ & $ \{u_1 \}$ & $ \{ u_2 \}$ & $  \{ [\{-2x_1+u_2, u_2-2x_2 \},1] \}$ \\
        \hline
        $\;\;3\;\;$ & $ \{u_1-u_2 \}$ & $ \{u_2 \}$ &  $ \{ [\{x_1, x_2-u_2\},1], [\{2x_1-3u_2, x_2+u_2\},1]\}$ \\
        \hline
        $\;\;4\;\;$ & $ \{u_1+u_2 \}$ &  $ \{u_2 \}$ & $\{[\{x_1, x_2-u_2\},1],  [\{2x_1-3u_2, 2x_2+u_2\},1]\}$\\
        \hline
        $\;\;5\;\;$ & $ \{u_2 \}$ & $\{u_1\}$ & $ \{[\{ 2x_1^2-u_1^2, 2x_2^2-u_1^2\},1]\}$ \\
        \hline
        $\;\;6\;\;$ & $\{u_1,u_2\}$  & $\{1\}$ &  $\{ [\{x_2\},1],  [\{x_1, x_2\},1], [\{2x_1^2-u_1^2,2x_2^2-u_1^2\},1]\}$\\
         \hline
      \end{tabular}
  \end{center}
     \end{Table}

%

\end{Example}





\section{Experiment of Comparison}\label{scwc}
We have implemented the Algorithm {\tt HCTD} as a  Maple function ${\tt HCTD}$ and tested a great many  benchmarks from the references \cite{zxq,changbo,sun,Montes}.
Throughout this paper,
 all the computational results are obtained in Maple 17 using an Intel(R) Core(TM) i5 processor (3.20GHz CPU), 2.5 GB RAM and Windows 7 (32 bit).   All the timings are given by seconds. The ``{\tt timeout}" mark means the time is greater than 1000 seconds.
The Table \ref{res1} compares the functions {\tt HCTD} (when $m=d$) and {\tt ComprehensiveTriangularize} ({\tt CTD}) in {\tt RegularChains}.

In Table \ref{res1}, the column ``time" lists the timings of $\tt{HCTD}$ ($m=d$) and $\tt{CTD}$;
the column ``branch" lists the numbers of branches output by $\tt{HCTD}$ and $\tt{CTD}$; and
the column ``grape" lists the numbers of grapes output by $\tt{HCTD}$ and $\tt{CTD}$.
It is indicated by  Table \ref{res1} that
   \begin{itemize}
      \item  
   for the benchmarks 3--27, {\tt HCTD} runs  faster than {\tt CTD}, especially, for the benchmark 27, {\tt CTD} is timeout and
  {\tt HCTD} completes the computation in time;
  for the benchmarks 28--40, {\tt CTD} runs faster than {\tt HCTD}, especially, for the benchmarks 38--40, {\tt HCTD} is timeout and
  {\tt CTD} solves the systems efficiently;
  for the benchmarks 41--49, both {\tt HCTD} and {\tt CTD} are timeout;
 \item  for the benchmarks 14, 31, 32, 35 and 36, the number of  branches output by {\tt HCTD} is much bigger than that output by {\tt CTD};
 \item  for the benchmarks 6, 10, 12, 29, 30, 32, 35 and 37, the number of  grapes output by {\tt HCTD} is much bigger than
 that output by {\tt CTD}.
  \end{itemize}
\section{Different Hierarchical Strategy}\label{sva}
To compute a $m$-HCTD for a given parametric system, as shown by Example \ref{ex1}, we first take $\{x_1,\ldots,x_n\}$ as variable set and then add one parameter into the variable set at each recursive step. A different hierarchical strategy may be that we add a prescribed number (say $s$) of parameters into the variable set at the first step and each recursive step.

The algorithm applying this different hierarchical strategy is called {\tt HCTDA} and has been implemented as a function {\tt HCTDA}.
The comparing data of {\tt HCTD} and {\tt HCTDA} (for $s=1$) is shown in Table \ref{res2}. It is indicated by  Table \ref{res2} that                                                                                                               \begin{itemize}
      \item  
   for the benchmarks 3--11, {\tt HCTD} runs  faster than {\tt HCTDA}, especially, for the benchmarks 10--11, {\tt HCTDA} is timeout
      and
  {\tt HCTD} completes the computation in time;
  for the benchmarks 12--18, {\tt HCTDA} runs faster than {\tt HCTD}, especially, for the benchmarks 19--20, {\tt HCTD} is
      timeout and
  {\tt HCTDA} completes the computation in time;

 \item  the difference of the numbers of branches (grapes) output by {\tt HCTD} and {\tt HCTDA} is not striking.
 \end{itemize}
In fact, we can input different $s$ when calling {\tt HCTDA}. For many benchmarks in Table \ref{res1}, the timings of different $s$ are similar. There are some benchmarks on which the timings of {\tt HCTDA} differ greatly  for different $s$. Due to page limitation, we do not report the timings here.

\section{Benefit of Hierarchical strategy}\label{sps}

We see that the benchmarks 41--49 in Table \ref{res1} are timeout when using both {\tt CTD} and {\tt HCTD} ($m=d$). In fact, for some polynomial systems from practical areas,  the complexity of
computing comprehensive triangular decomposition is way beyond current computing
capabilities. However for these systems (especially the systems with many parameters), we may try to compute the $m$-HCTD for
$m=0,\ldots,d-1$. In this way,  although we cannot solve  the system completely, we may still get partial solutions.

We have tried the {\tt timeout} benchmarks 41-49 in Table \ref{res1}. The experimental results are shown in Table \ref{res4}, where the columns ``$m=0$", ``$m=1$", ``$m=2$", ``$m=3$" and ``$m=4$" denote the timings of calling Algorithm {\tt HCTD} for $m=0, \;1,\; 2,\; 3,\; 4$; and the ``{\tt error}"  mark means  Maple returns an error message and stops computing. It is seen from the Table \ref{res4} that
\begin{itemize}
\item for all the benchmarks, we successfully get partial solutions;
\item for most of the benchmarks, such as the benchmark 1 and benchmarks 3--7, we get results only when $m=0$. 
\end{itemize}

\begin{Table}\label{res1}\footnotesize
 \begin{center}
      {Comparing {\tt HCTD} and \tt{CTD}}\\
      \begin{tabular}{|c|c|c|c|c|c|c|c|c|c|}
      \hline
        & \bbb{benchmark} &\bbb{$d$} &\bbb{$n$} & \multicolumn{2}{|c|}{time}  &
       \multicolumn{2}{|c|}{branch } & \multicolumn{2}{|c|}{grape } \\
      \cline{5-10}
        & & & & ${\tt HCTD}$ & ${\tt CTD}$ & ${\tt HCTD}$ & ${\tt CTD}$ &${\tt HCTD}$ & ${\tt CTD}$\\
        \hline
       1.  &{\em MontesS2} &1&3& 0.&	0.&	1&	1&	1&	1\\
       2.  &{\em MontesS4} &2&2& 0.	&0.	&1	&1	&1&	1\\
       3.   &{\em F8}&4&4&	0.437&	1.014&	18&	14	&14	&9\\
        4.  &{\em Hereman-2} &1&7 &0.093&	0.468	&2&	2&	10&	6\\
        5. &{\em MontesS3} &1&2& 0.	&0.031&	3&	2	&2&	2\\
        6.  &{\em MontesS5}&4&4&	0.078&	0.187&	6&	8	&13	&6\\
        7.  &{\em MontesS6}&2&2&0.015&	0.047&	4&	3&	5&	4\\
       8.  & {\em  MontesS7}&	1&	3&	0.046&	0.156&	4&	4&	6&	8\\
       9.  &{\em MontesS8}&2&2&0.&	0.094&	2&	2&	2&	2\\
        10.  &{\em MontesS12}&2&6&0.593&	7.925&	5&	5&	61&	27\\
       11. &{\em MontesS13} &3&2&0.078&	0.265&	6&	9	&9&	8	\\
       12.  &{\em MontesS14}&1&4&0.452&	4.353&	6&	3&	28&	12\\
       13. &{\em MontesS15}&4&8&0.187&	0.889&	5&	5&	14&	12\\
       14.  &{\em MontesS16}&3&12&1.198&	1.825&	37&	8&	11&	7\\
       15.   &{\em Bronstein}&2&2 &0.015 &	0.219&	6&	7&	7&	7\\
       16.  &{\em AlkashiSinus}&3&6 &0.094&	0.437&	8&	6&	8&	6\\
        17. &{\em Lanconelli}&7&4&0.28&	0.546&	14&	11&	7&	5\\
         18.  &{\em zhou1}  &3&4&0.047&	0.156&	5&	5&	5	&5\\
         19.  &{\em zhou2}  &6&7&0.671&	2.09&	17&	18&	19&	16\\
         20.  &{\em zhou6}  &3&3&0.031&	0.218&	6&	4&	6&	5 \\
         21.  &{\em SBCD13}&1&3&	0.015&	0.094	&2&	2&	9&	6\\
         22. &{\em SBCD23}&1&3&	0.202	&0.344&	4	&2&	15&	12\\
     23.   &{\em F2}&2&2&	0.032	&0.234	&3&	3&	3&	3\\
     24.  &{\em F3}&4&1&	0.063&	0.905&	5&	6&	5	&6\\
     25. &{\em F5}&3&2&	0.046&	0.11&	6&	3&	3&	3\\
     26.    &{\em F7}&3&2&	0.&	0.016&	2&	2&	2&	2\\
     27.  &{\em S2}&4&1&	44.544	&{\tt timeout}&	150& &	92&	\\
     28. &{\em MontesS9}&3&3&0.693&	0.468&	21&	13&	16	&13\\
     29. &{\em MontesS10}&3&4&0.421&	0.359&	13&	6&	19&	6 \\
     30. &{\em MontesS11}&3&3&0.858&	0.655&	12&	16&	20&	10\\
     31.&{\em F4}&4&2&	11.637	&0.375&	20&	3&	3&	3\\
     32. &{\em zhou5}  &4&5&5.616&	2.902&	51&	19&	97&	22\\
     33.&{\em F6}&4&1&	0.296&	0.14&	13	&3&	11&	3\\
     34. &{\em MontesS1} &2&2& 0.016&	0.&	4&	2&	3&	3\\
     35. &{\em Hereman-8-8}  &3&5&96.439&	10.468&	108&	9&	161&	14 \\
     36. &{\em S3}&4&3&	2.618&	1.436&	35&	13&	17&	11\\	
     37. &{\em Maclane} &3&7& 5.242&	4.009&	17&	9&	155&	27\\
     38.&{\em S1}&3&2&	{\tt timeout}&	4.04&	&	10&	&	10\\
     39.&{\em Neural}&1&3&{\tt timeout}	&0.188&	&	2& &		7\\	
     40.&{\em Gerdt}	&3	&4	&{\tt timeout}&	0.842	&	&4&	&	6\\
     41.&{\em Lazard-ascm2001}&	3	&4&	{\tt timeout}&{\tt timeout}& & & & \\			
     42.&{\em Leykin-1}&	4&	4	&{\tt timeout}&{\tt timeout}& & & & \\		
     43.&{\em Cheaters-homotopy-easy}&	4&	3	&{\tt timeout}&{\tt timeout}& & & &  \\			
     44.&{\em Cheaters-homotopy-hard}&	5&	2&	{\tt timeout}&{\tt timeout}& & & & \\			
     45.&{\em Lazard-ascm2001}&	3	&4		&{\tt timeout}&{\tt timeout}& & & &  \\		
     46.&{\em MontesS18}&	2&	3	&{\tt timeout}&{\tt timeout}& & & &  \\			
     47.&{\em Pavelle} &	4&	4	&{\tt timeout}&{\tt timeout}& & & &  \\		
     48.&{\em p3p} &	5&	2&	{\tt timeout}&{\tt timeout}& & & & \\
     49.&{\em z3}&	6&	11&	{\tt timeout}&{\tt timeout}& & & &  \\		
      \hline
     \end{tabular}
 \end{center}
    \end{Table}

  \begin{Table}\label{res2}\footnotesize
  \begin{center}
       {Comparing {\tt HCTD} and \tt{HCTDA}} (for $s=1$)\\
       \begin{tabular}{|c|c|c|c|c|c|c|c|c|c|}
       \hline
         & \bbb{benchmark} &\bbb{$d$} &\bbb{$n$} & \multicolumn{2}{|c|}{time}  &
        \multicolumn{2}{|c|}{branch } & \multicolumn{2}{|c|}{grape } \\
       \cline{5-10}
         & & & & ${\tt HCTD}$ & ${\tt HCTDA}$ & ${\tt HCTD}$ & ${\tt HCTDA}$ &${\tt HCTD}$ & ${\tt HCTDA}$\\
         \hline
        1. &{\em MontesS5}&4&4&	0.078&	0.078&	6&	6	&13	&13\\
        2.   &{\em zhou1}  &3&4&0.047&	0.047&	5&	5	&5	&6\\
        3.  &{\em MontesS9}&3&3&0.693&	0.796&	21&	21&	16&	27\\
        4. &{\em MontesS11}&3&3&0.858&	1.207&	12&	24&	20&	38\\
        5. &{\em MontesS12}&2&6&0.593&	0.671&	5&	5&	61&	60\\
        6.  &{\em AlkashiSinus}&3&6&0.094&	0.109&	8&	10&	8	&10\\
        7.   &{\em Bronstein}&2&2 &0.015&	0.031&	6&	5&	7&	6\\
        8.  &{\em MontesS7}&2&2&0.046&	0.266&	4&	6&	4&	6\\
         9.   &{\em SBCD13}&1&3&0.015&	0.031&	2&	2&	9&	7\\
          10. &{\em F6}&4&1&	0.296&{\tt	timeout}&	13& &		11&	\\
          11.  &{\em S2}&4&1&	44.544&{\tt	timeout}&	150& &		92&	\\
          12. &{\em Maclane} &3&7& 5.242	&2.605&	17&	13&	155&	122\\
      13.&{\em SBCD23}&1&3&	0.202&	0.109&	4&	2	&15	&13\\
      14.&{\em F4}&4&2&	11.637&	1.653&	20&	26&	3&	3\\
      15.  &{\em MontesS15}&4&8&0.187&	0.124&	5&	5&	14&	14\\
      16.&{\em F8}&4&4&	0.437&	0.358	&18&	16&	14&	11\\
      17. &{\em MontesS16}&3&12&1.198&	0.951&	37&	21&	11&	8\\
      18.&{\em S3}&4&3&	2.618&	1.81	&35&	29&	17	&15\\	
      19.&{\em Neural}&1&3&{\tt timeout}&	0.296&  &		6	& &	15\\	
      20.&{\em Gerdt}	&3	&4	&{\tt timeout}&	288.352	 & &	4	& &	11\\
       \hline
      \end{tabular}
  \end{center}
     \end{Table}

  \begin{Table}\label{res4}\footnotesize \begin{center}
     {Timings of $m$-HCTD for different $m$}\\
     \begin{tabular}{|c|c|c|c|c|c|c|c|c|}
                \hline
       & \bbb{benchmark} &\bbb{$d$} &\bbb{$n$}  & \multicolumn{5}{|c|}{time}  \\
     \cline{5-9}
        & & &   &$m=0$& $m=1$ & $m=2$& $m=3$ & $m=4$\\
       \hline
        1.&{\em Lazard-ascm2001}&	3	&4&	0.936&{\tt timeout}& & &\\			
        2.&{\em Leykin-1}&	4&	4	&0.203&	20.436&{\tt timeout} & &\\		
        3.&{\em Cheaters-homotopy-easy}&	4&	3	&3.681&{\tt timeout} &  & &\\			
        4.&{\em Cheaters-homotopy-hard}&	5&	2&	39.640 &{\tt timeout} & & &\\			
        5.&{\em Lazard-ascm2001}&	3	&4		&0.858&{\tt timeout} &  & &\\		
        6.&{\em MontesS18}&	2&	3	&0.327&{\tt timeout} &  & &\\			
        7.&{\em Pavelle} &	4&	4	&0.234&{\tt timeout} 	& & &\\		
        8.&{\em p3p} &	5&	2&	0.&	0.&	0.015	&6.549&{\tt timeout}\\
        9.&{\em z3}&	6&	11&	0.094&	{\tt error} &  & &\\		
     \hline
    \end{tabular}
\end{center}
   \end{Table}

\end{document}